# X-ray polarimetry with an active-matrix pixel proportional counter

J.K. Black[a,*], P. Deines-Jones[b], S.E. Ready[c], R.A. Street[c]

[a] *Forbin Scientific, Code 662, NASA/Goddard Space Flight Center, Greenbelt, MD, 20771, USA*

[b] *Universities Space Research Association, Code 661, NASA/Goddard Space Flight Center, Greenbelt, MD, 20771, USA*

[c] *Palo Alto Research Center, 3333 Coyote Hill Road, Palo Alto, CA, 94304 USA*

**Abstract**

We report the first results from an X-ray polarimeter with a micropattern gas proportional counter using an amorphous silicon active matrix readout. With 100% polarized X-rays at 4.5 keV, we obtain a modulation factor of 0.33 ± 0.03, confirming previous reports of the high polarization sensitivity of a finely segmented pixel proportional counter. The detector described here has a geometry suitable for the focal plane of an astronomical X-ray telescope. Amorphous silicon readout technology will enable additional extensions and improvements.

The prospects for X-ray polarimetry as a practical tool have been revolutionized by the recent introduction of an instrument with unprecedented sensitivity [1-3]. Based on the photoelectric effect, this new type of polarimeter uses a finely spaced, gas pixel detector to image the tracks of photoelectrons. Polarization information is obtained by determining the angle of emission of the photoelectron, which is correlated with the electric field vector of the X-ray.

The primary innovation of this new polarimeter is the introduction of a pixel readout anode to a micro-pattern gas proportional counter. With a pixel size small compared to the photoelectron range, photoelectron tracks can be imaged and the emission angle reconstructed event-by-event. Combined with the high quantum efficiency and broad bandpass of the photoelectric effect, this non-dispersive technique gives dramatic improvements in sensitivity over polarimeters based on X-ray scattering [4].

Further improvements in the pixel anode will allow this new polarimetry technique to realize its full potential. The polarimeters demonstrated previously consist of a gas electron multiplier (GEM) [5] suspended above a readout plane segmented into individual pixels. Each pixel is connected to a chain of readout electronics through a multi-layer fine-line printed circuit. This type of pixel anode limits both the pixel spacing and the total number of pixels. An anode plane with integrated active electronics, such as CMOS VLSI [3], would overcome this limitation.

Another possible active pixel anode plane is an amorphous silicon thin-film transistor (TFT) array like those used in flat-panel imagers [6]. In this active-matrix addressing scheme, each pixel is connected to readout electronics through a thin-film transistor that acts as an analog switch. Charge is held on the pixel until the transistor is activated, allowing a multiplexed readout such that $n^2$ pixels are read out with $2n$ electronic channels.

We report the first results from a polarimeter with TFT active-matrix readout and a geometric area suitable for the focal plane of a typical X-ray astronomical telescope with conical foil mirrors [7,8].

---

[*] Corresponding author. Tel.: +1-301-286-1231; fax: +1-301-286-1684; e-mail: black@forbinsci.com.

Photoelectron track images were recorded from X-rays of energy 4.5, 5.9 and 20 keV. Polarization measurements were made with 4.5 keV X-rays.

The detector, shown schematically in Fig. 1, consisted of a double-GEM with 100-μm hexagonal pitch and a TFT pixel anode with 100-μm square pitch. The active area was defined by the double-GEM, which was 12mm x 12mm. Only a small fraction of the TFT array's 512 x 512 pixels (about 25 $cm^2$) was used. No attempt was made to align the GEMs with each other or with the TFT array.

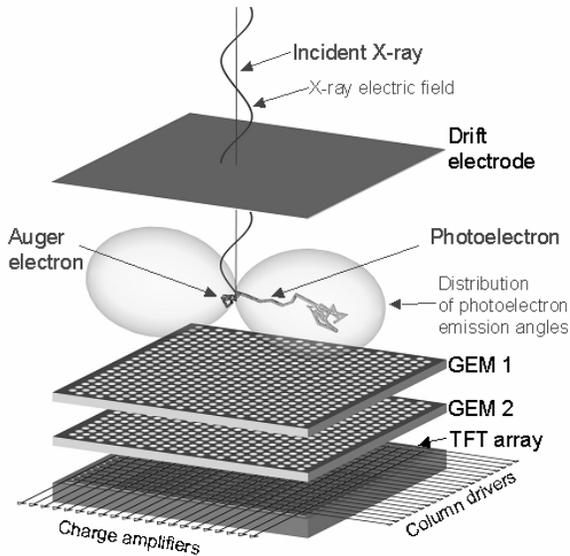

Fig. 1. Schematic diagram of detector geometry used in these measurements. The $\sin^2\theta\cos^2\phi$ distribution of photoelectron emission for normally incident X-rays is projected onto the detector plane and observed as $\cos^2\phi$.

The GEMs were contained in a detector body with a thin polyimide entrance window. The active depth was 3mm, determined by the spacing between a nickel mesh drift electrode and the GEMs. The spacing between GEMs was 750 μm. The spacing between the GEMs and the TFT array was maintained by a 750 μm thick silicone gasket that also acted as a gas seal. The detector was operated as a flow counter using an 80% neon, 20% $CO_2$ gas mixture at an apparent gain of about $10^4$. The drift field was 300 V/cm.

The GEMs were fabricated by UV laser ablation [9,10] in the Laboratory for High Energy Astrophysics at NASA's Goddard Space Flight Center (GSFC). The GEMs are 75 μm thick with hole diameters that are 60 μm at the electrodes, constricting to 50 μm in the center of the substrate.

The TFT pixel arrays were fabricated at the Palo Alto Research Center using a process scalable to large areas (~$10^3$ $cm^2$). Each pixel contains a metal anode pad, a storage capacitor of about 0.4 pF and a TFT switch. The pixel design is described in more detail elsewhere [11,12]. The source of the TFT is connected to the anode pad, which is isolated from the rest of the pixel by an insulation layer. The anode pads cover about 67% of the surface area. The TFT gates are connected in columns and the drains are connected in rows.

With the TFT switches open, charge from the GEMs is stored on the anode pads. To read out the array, the TFT switches are closed column-by-column by activating the gates lines, so that charge is transferred from a column of anode pads to a row of external amplifiers [6]. Typical operating conditions turn on the TFT gate for 20–30 microseconds, which is nearly 10 times the *RC* time constant of the TFT resistance and pixel capacitance. The signal is digitized to 14 bits using correlated double sampling. Electronic noise is less than 1000 electrons (rms). For these measurements, the array was operated in a scanning mode at a frequency of 6 Hz.

Fig. 2 shows a sample of tracks recorded from 5.9 keV X-rays. The non-uniform ionization density reveals details about the X-ray interaction. Most of the charge is deposited at the end of the photoelectron track. At the beginning of the track is a smaller cluster of charge deposited by the Auger electron (870 eV in neon), which is emitted isotropically.

Significantly more structure is seen in tracks from 20 keV photons as shown in Fig. 3. The Auger/interaction point can still be identified among more hard scatters and knock-on electrons. These tracks emphasize that large-area detectors of this type could be used to track high-energy electrons or minimum ionizing particles.

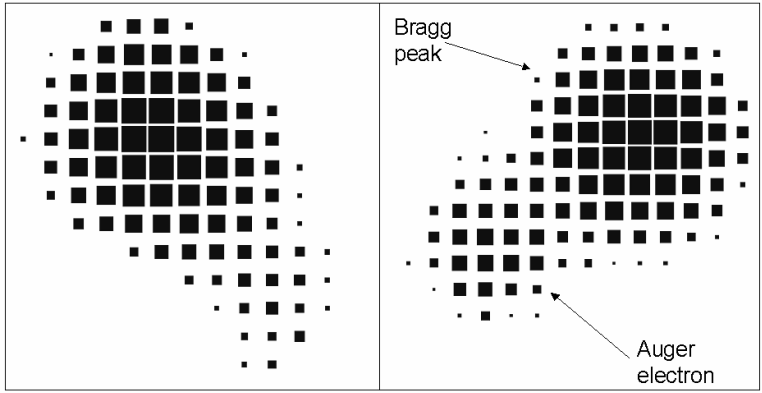

Fig. 2. Example track images from 5.9 keV X-rays. The boxes are on a 100 micron grid. The area of each box represents the total charge collected in the pixel. Two peaks are visible in the distribution. The smaller peak is due to the Auger electron, emitted at the X-ray interaction point.

At 4.5 keV, most of this fine structure is obscured by the diffusion of the primary electrons in the drift volume as shown in Figure 3. Nevertheless, the track images have a discernable direction.

Polarized photons were produced by reflecting X-rays at 90 degrees off a silicon crystal. The X-rays were produced by a tube with a titanium anode ($K_\alpha$ = 4.5 keV). Data were collected with the plane of X-ray polarization at three different angles, separated by about 45 degrees, with respect to the detector.

The polarized data were analyzed by reconstructing the emission angle of each photoelectron. Histograms of the emission angles were then fit to the expected functional form: $N(\phi) = A + B\cos^2(\phi - \phi_0)$, where $\phi_0$ is the angle of the plane of polarization [13]. The sensitivity to polarization is defined by the modulation:

$\mu = (N_{max} - N_{min})/(N_{max} + N_{min}) = B/(2A + B)$,

where $N_{max}$ and $N_{min}$ are the maximum and minimum of the function, respectively.

The photoelectron emission direction of each event was determined by calculating the direction of the major axis of the second moment of the charge distribution about its barycenter. Only pixels above a threshold of 2000 electrons were included in the calculation. A more refined analysis, including only the charge near the interaction point [13], was not attempted.

The results shown, in Fig. 4, demonstrate the polarization sensitivity of the detector. The data fit the expected functional form and peak around the independently measured polarization angle of the incident X-rays. All three data sets are consistent with the average modulation of 0.33 ± 0.03, which is also consistent with results from simulated data. The fit parameters are given in Table 1.

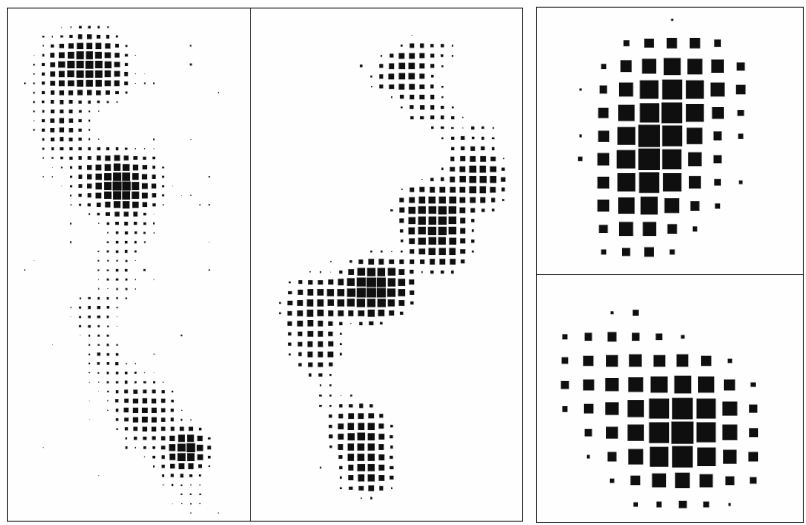

Fig. 3. Track images from 20 keV X-rays (left) and 4.5 keV X-rays (right).

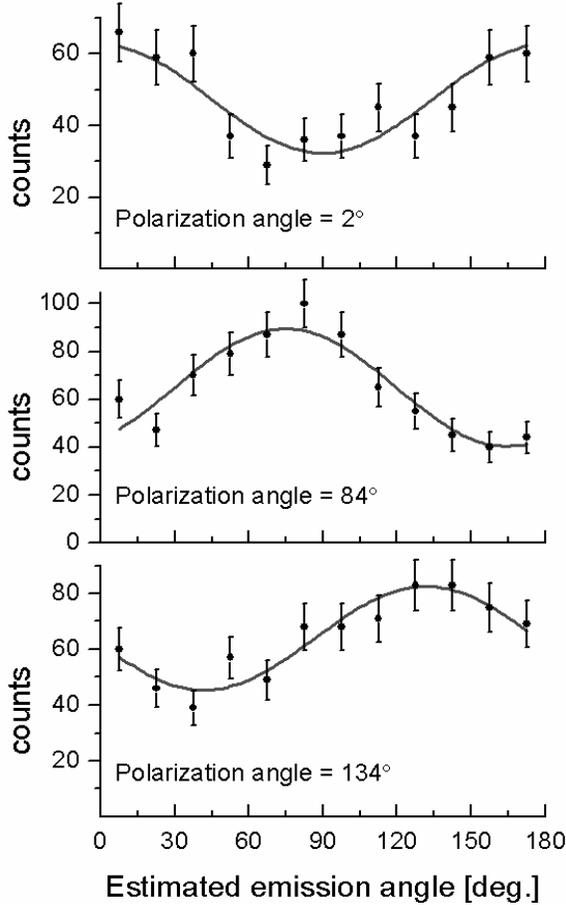

Fig. 4. Histograms of reconstructed emission angles for three polarization angles. The curves are fits to the data.

|  | Fit Parameters | |
| --- | --- | --- |
| Polarization Angle | $\mu$ | $\phi_0$ |
| 2 | 0.30 ± 0.06 | 0 ± 5 |
| 84 | 0.38 ± 0.05 | 75 ± 4 |
| 134 | 0.29 ± 0.05 | 132 ± 5 |

Table 1. Fit results to reconstructed emission angles. Stated errors are one standard deviation.

We have demonstrated a micropattern photoelectric polarimeter with a geometry suitable for the focal plane of an astronomical X-ray telescope. Amorphous silicon TFT anodes offer numerous possibilities for further improvement and extension. TFT arrays can be fabricated in large areas, opening the possibility of large-area collimated detectors, which might be a more practical alternative to an X-ray optic in a small satellite mission. Amorphous silicon can also be fabricated on a variety of optically thin substrates, including thin plastics [14]. Optically thin detectors would allow the stacking of multiple polarimeters at the focus of an X-ray mirror, with each layer optimized for a higher energy than the one before. In this way, the quantum efficiency and bandpass of the instrument could be even further extended.

## Acknowledgements

This work was funded by NASA grant number NRA-01-01-HEA-021 and the GSFC Director's Discretionary Fund. Numerous individuals in the Laboratory for High Energy Astrophysics at GSFC supported this work. We particularly thank Mike Lenz, Bert Nahory, Norman Dobson and Dr. Scott Owens for crucial technical support and Dr. Keith Jahoda for many helpful discussions. We also thank the members of the PARC process line for fabricating the TFT array.